\documentclass[aps,floatfix,prc,twocolumn,showpacs,showkeys,amsmath,amssymb,superscriptaddress]{revtex4}
\usepackage{graphicx}
\usepackage{dcolumn}% Align table columns on decimal point
\usepackage{bm}% bold math
\usepackage{color}
\def\pp{p+p~}
\def\sr17{$\sqrt{s}$~=~17~GeV~}
\newcommand{\pbar}{$\rm\overline{p}$}
\newcommand{\Bbar}{$\rm\overline{B}$}
\newcommand{\Lbar}{$\rm\overline{\Lambda}$}
\newcommand{\La}{$\Lambda$}

\newcommand{\sNN}{$\sqrt{s_{\rm NN}}$}
\newcommand{\be}{\begin{equation}}
\newcommand{\ee}{\end{equation}}

\def\muB{$\mu_B$}
\def\muS{$\mu_S$}
\begin{document}
\title{Antimatter production in proton-proton and heavy-ion collisions at ultrarelativistic energies}
\author{J.~Cleymans}
\affiliation{UCT-CERN Research Centre and Department  of  Physics,\\ University of Cape Town, Rondebosch 7701, South Africa}
\author{S.~Kabana}
\affiliation{SUBATECH, 4 rue Alfred Kastler, F-44307 Nantes,
France}
\author{I.~Kraus}
\affiliation{Institut f\"ur Kernphysik, Darmstadt University of
Technology, D-64289 Darmstadt, Germany}
\author{H.~Oeschler}
\affiliation{Institut f\"ur Kernphysik, Darmstadt University of
Technology, D-64289 Darmstadt, Germany} \affiliation{European
Organization for Nuclear Research (CERN), Geneva, Switzerland}
\author{K.~Redlich}
\affiliation{Institute of Theoretical Physics, University of Wroc\l aw, Pl-45204 Wroc\l aw, Poland}
\affiliation{ExtreMe Matter Institute EMMI, GSI, D-64291 Darmstadt, Germany}
\author{N.~Sharma}
%\affiliation{European Organization for Nuclear Research (CERN), Geneva, Switzerland}
\affiliation{Panjab University, Chandigarh, India}
%\affiliation{GSI Hemholtzzentrum f\"ur Schwerionenforschung, D-64291 Darmstadt, Germany}
%\affiliation{ExtreMe Matter Institue EMMI, GSI, D-64291 Darmstadt, Germany}

\date{\today}
\begin{abstract}
One of the striking features of particle production at high beam
energies is the near equal abundance of matter and antimatter in
the central rapidity region. In this paper we study how this
symmetry is reached as the beam energy is increased. In
particular, we quantify explicitly the energy dependence of the
approach to matter/antimatter symmetry  in proton-proton and in
heavy-ion collisions. Expectations are presented also for the
production of more complex forms of antimatter like
antihypernuclei.

\end{abstract}
\maketitle
\section{\label{secIntroduction}Introduction}
%
% ----------- antimatter

 One of the striking features of particle production at high
energies is the nearly equal abundance of matter and antimatter in
the central rapidity region~\cite{Abelev:2008ez,Aamodt:2010d}. It
is believed that a similar symmetry existed in the initial stage
of the universe. It  remains a mystery  how this symmetry got lost
in the evolution of the universe reaching a stage with no visible
amounts of antimatter being present. Closely related to the
matter-antimatter symmetry is the production of light antinuclei,
hypernuclei and antihypernuclei at high energies, especially in
view of the recent observation of the anti $^4$He nucleus by the
STAR Collaboration~\cite{Agakishiev:2011ib}.

Since the first observation of hypernuclei in 1952
\cite{Danysz:1953aa} there has been a steady interest in searching
for new hypernuclei and exploring the hyperon-nucleon interaction
which is relevant (see e.g.~\cite{Hahn:1986mb,Stoecker:1986ci})
for nuclear physics. Hypernuclei decay with lifetime which depends
on the strength of the hyperon-nucleon interaction. While several
hypernuclei have been found since the first observation no
antihypernucleus has ever been observed until the recent discovery
of the antihypertriton in Au+Au collisions at $\sqrt{s_{\rm NN}}$
= 200 GeV by the STAR Collaboration at RHIC \cite{Abelev:2010}.
The yield of (anti)hypernuclei  measured  by STAR is very large,
in particular they seem to be produced with a similar yield as
other (anti)nuclei like e.g.~(anti)$^3$He. This abundance is much
higher than measured for hypernuclei and nuclei at lower
energies~\cite{Rapp:2000gy}. It is of interest to understand the
nature of this enhancement, and for this, the systematics of
antimatter production in high energy hadron-hadron and heavy-ion
collisions should be investigated.

The analysis of particle production assessing the degree of
thermalization of the particle source has been undertaken since
many decades
\cite{Fermi:1950jd,Pomeranchuk:1951ey,Heisenberg:1952zz,Landau:1953gs,Hagedorn:1965st,Hagedorn:1968zz}.
It has been found that the thermalization assumption applies
successfully to hadrons produced in a large number of particle and
nuclear reactions at different energies
\cite{BraunMunzinger:2003zd,Redlich:2002ij,BraunMunzinger:2001as}.
This fact allowed  to estimate thermal parameters characterizing
the particle source for each colliding system which is    relevant
for the understanding of the thermal properties of dense and hot
matter and for studies of QCD phase transitions
\cite{BraunMunzinger:1996mq,Karsch:2010ck}.

In this paper,
 using the energy dependence of thermal parameters obtained from  the statistical thermal model
analysis of particle yields in heavy-ion collisions
~\cite{Wheaton:2004vg,Wheaton:2004qb} we present the  model
estimates for the  (anti)hypernuclei multiplicity  that can be
directly compared to the recent results obtained in central Au+Au
collisions at RHIC.   We  discuss systematics of (anti)matter
production  at different   energies. We also make  predictions of
(anti)matter and (anti)hypernuclei production at the Large Hadron
Collider (LHC).

Recently, a very interesting  analysis of  the production of light
nuclei, hypernuclei and their antiparticles in central heavy-ion
collisions was  performed in Ref.~\cite{Andronic:2010qu} within
the statistical thermal model. It was found that ratios of
hypernuclei to Lambda exhibit an energy dependence similar to the
K$^+/\pi^+$ ratio with a clear maximum at low energy. The present
work is considered  to be an extension of the analysis performed
in Ref.~\cite{Andronic:2010qu}.

Firstly,  we compare the statistical thermal model results  on
(anti)baryon production in heavy-ion and in proton-proton
collisions. This, in general,  requires   the knowledge  of  the
energy dependence of thermal  parameters in \pp collisions which
are proposed in this paper based on the recent data. In this
context,  we study quantitatively how the matter/antimatter
symmetry is reached as the beam energy is increased both for \pp
and heavy-ion collisions. We also discuss the role of the
strangeness content of particles and quantify different
antibaryon/baryon ratios in \pp and in heavy-ion collisions at
SPS, RHIC and LHC energies.

Secondly, we compare predictions of the statistical thermal and
coalescence models for different ratios of (anti)nuclei and
(anti)hypernuclei in the context of recent STAR data obtained in
central Au+Au collisions at the top RHIC energy.

The paper is organized as follows: In section II we discuss
features of the statistical thermal model.  In section III we
compare the antibaryon/baryon ratios in \pp and heavy-ion
collisions and obtain  the energy dependence of thermal parameters
in \pp collisions. We demonstrate the scaling behaviour of the
antibaryon/baryon ratio  with their  strangeness content. In
section IV we apply the thermal  and the coalescence models   to
the production of nuclei and hypernuclei and their antiparticles.
We also make predictions for (anti)nuclei and (anti)hypernuclei
yield ratios  at LHC energy.
In section V we summarize  our results.\\

\section{\label{model}The statistical thermal model}

The statistical thermal model assumes that in a high energy
collision at freeze-out all hadrons  follow equilibrium
distributions. The conditions at chemical freeze-out where
inelastic collisions cease are given by the hadron abundances,
while the particle spectra offer insight into the conditions at
thermal freeze-out where elastic collisions cease. Once thermal
parameters are fixed, the hadron gas partition function gives all
primordial thermodynamic observables  of  the system.  The exact
form of the partition function, however, depends on the
statistical ensemble under consideration.

Within the grand-canonical (GC) ensemble, the quantum numbers of
the system are conserved on average through the action of chemical
potentials~\cite{Redlich:2002ij}. In other words, the baryon  $B$,
strangeness $S$ and the charge content $Q$ are fixed on average by
the $\mu_B$, $\mu_S$ and $\mu_Q$ chemical potentials respectively.
For each chemical potential one can introduce  the corresponding
fugacity $\lambda=e^{\mu/T}$ where $T$ is the temperature of the
system.

In the GC ensemble the  density of hadron species $i$ with the
mass $m_i$, the  quantum numbers $B_i$, $S_i$ and $Q_i$ and with
the spin-isospin degeneracy factor $g_i$ is express through the
second order modified Bessel function  $K_{2}(x) $ as
\begin{equation}
{n_i}(T,\mu_B,\mu_S,\mu_Q) =
\frac{g_i}{2\pi^2}m_i^2T\lambda_B^{B_i}\lambda_S^{S_i}\lambda_Q^{Q_i}
K_2(\frac{m_i}{T}) \label{eq:dens}.
\end{equation}
The above form, valid in the Boltzmann approximation,  is easily
generalized to the quantum
statistics~\cite{Wheaton:2004vg,Wheaton:2004qb}.

In the application of the statistical thermal model
%to particle production in collisions
the chemical potentials $\mu_S$ and $\mu_Q$ are typically
constrained in   the initial stage by the strangeness neutrality
condition and by the fixed baryon-to-charge ratio. However, it is
well established,  that the usual form of the statistical thermal
model formulated in the GC ensemble cannot be used when  either
the temperature or the volume parameter  $V$ or both
are small  \cite{Redlich:2002ij,Hamieh:2000tk}. %For charge neutral systems,
As a thumb rule one needs $VT^3>1$ for a grand canonical
description to
hold~\cite{Hagedorn:1984uy,Rafelski:1980gk,Cleymans:1998yb}. This
condition is not usually  justified  in \pp collisions,  requiring
canonical (C) formulation of strangeness conservation. The exact
strangeness conservation  causes a  suppression in particle ratios
of strange (or multi-strange) hadrons to pions or any strangeness
neutral particles as compared to the corresponding ratio in the
grand canonical limit. The key parameter governing this effect can
be quantified by the strangeness correlation
volume~\cite{Hamieh:2000tk}.

\section{\label{antibaryon}Production of antibaryons}

The chemical freeze-out conditions in heavy-ion collisions at
various energies were shown to follow a curve in the
temperature--baryo-chemical potential plane \cite{Cleymans:1998fq}
which has been phenomenologically parameterized as
~\cite{Cleymans:2005xv},
\begin{equation}
T(\mu_B) = a - b\mu_B^2 -c \mu_B^4  \label{Eqn:T}
\end{equation}
\begin{equation}
\mu_B = d / ( 1 + e \sqrt{s_{\rm NN}}) \label{Eqn:mub}
\end{equation}
with $ a =  0.166 \pm 0.002$ GeV, $b = 0.139 \pm 0.016$ GeV$^{-1}$
 $c = 0.053 \pm 0.021$ GeV$^{-3}$, $d = 1.308\pm0.028$ GeV
and $e = 0.273\pm0.008$ GeV$^{-1}$. This parametrization is
quantitatively similar to  the one proposed in
Ref.~\cite{Andronic:2005yp} and results in a very satisfactory
description of different particle excitation functions measured in
nucleus-nucleus   collisions.

\begin{figure}
\includegraphics[width=0.99\linewidth]{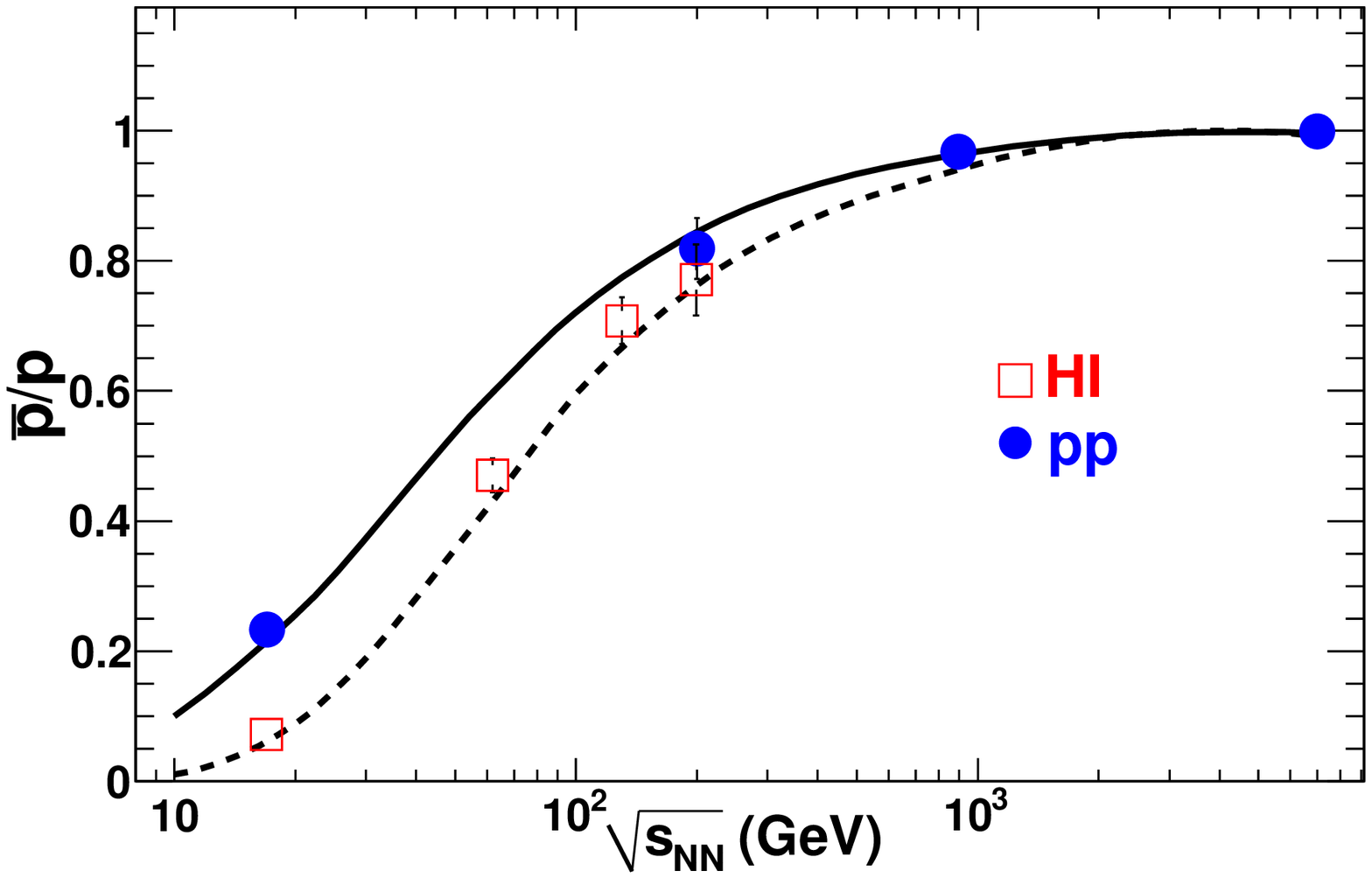}
\includegraphics[width=0.99\linewidth]{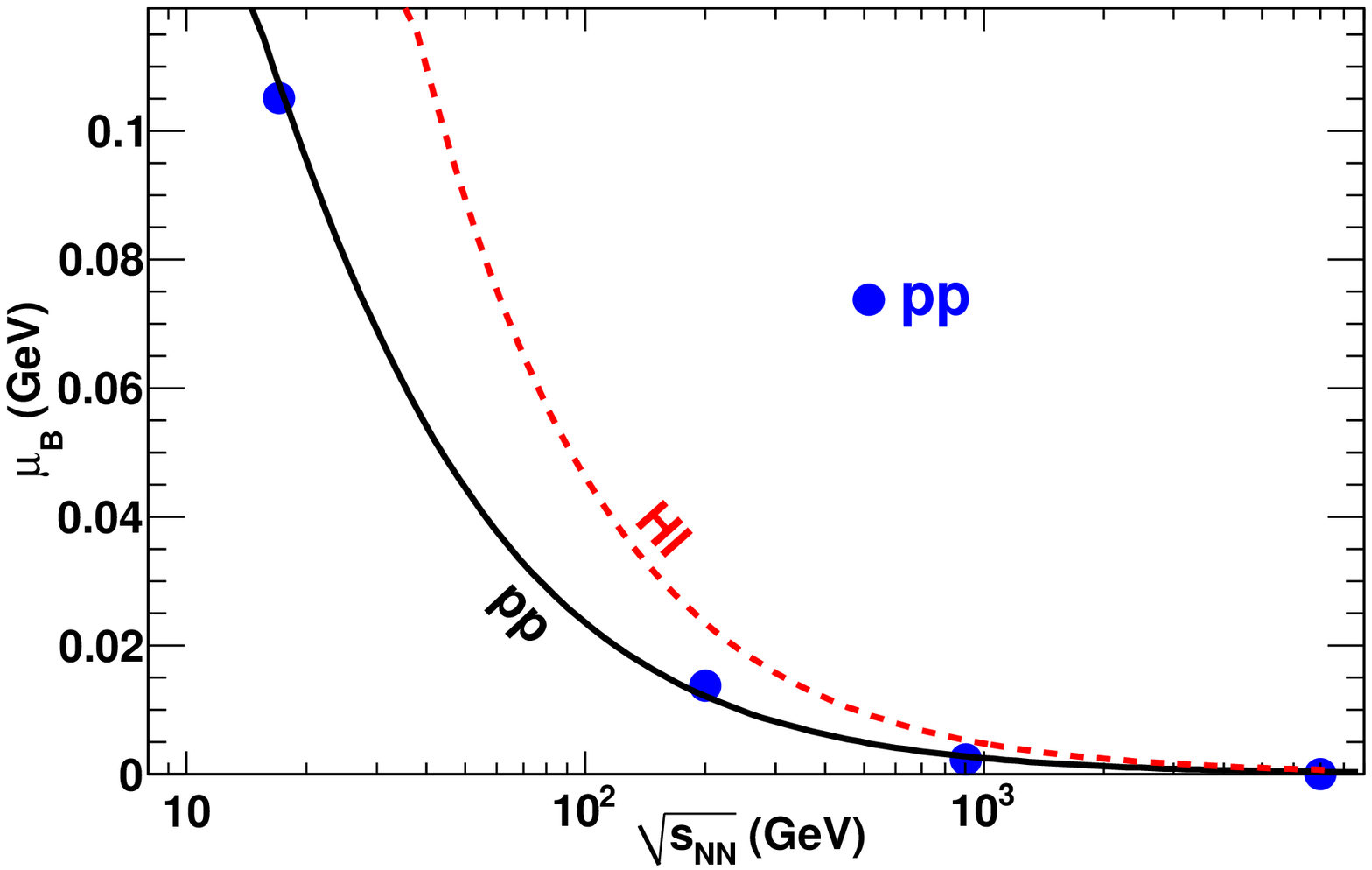}
\caption{The  \pbar/p ratio (upper part) and the corresponding
\muB (lower part) as a function of \sNN\ are shown. The solid
circles are results from \pp collisions and the open squares are
results from heavy-ion
collisions~\cite{Alt:2005gr,Abelev:2006cs,Aamodt:2010d,Alt:2007fe,Abelev:2008ez}.
The dashed line is the parametrization for heavy-ion collisions
from Ref.~\cite{Cleymans:2005xv} while the solid line is the new
parametrization for \pp collisions. } \label{pbarp}
\end{figure}

 The increase of  the  antimatter to matter
ratio in heavy-ion collisions with the center-of-mass energy  has been
observed  by the NA49~\cite{Alt:2005gr,Alt:2007fe} and the
STAR~\cite{Abelev:2006cs} Collaboration.
Fig.~\ref{pbarp} shows  changes of the
 \pbar/p
ratio  with collision  energy at mid-rapidity in  central
heavy-ion and  in  \pp collisions. The data  from NA49 and   STAR
Collaboration are compared with  new results from the ALICE
Collaboration~\cite{Aamodt:2010d}. There is a clear increase of
this ratio towards unity,   indicating approximate  symmetry of
matter and antimatter at the LHC  energy. There is also a clear
increase of  the \pbar/p ratio when going from  heavy-ion towards
\pp collisions.

In  Fig.~\ref{pbarp} data are compared with statistical thermal
model results. In heavy-ion collisions these model calculations
were done using the energy dependence  of model parameters as
described by Eqs.~(\ref{Eqn:T}) and (\ref{Eqn:mub}). There is a
clear agreement of model predictions with data.  For \pp
collisions no systematic analysis of model parameters with energy
were performed till now.

If (anti)nucleons are directly originating from a thermal source,
then from Eq.~(\ref{eq:dens}) (i.e. neglecting feed-down from resonances)
 it is obvious, that  the \pbar/p
densities ratio
\begin{equation}
\frac{n_{\rm\overline{p}}}{n_{\rm p}} =  \exp[{-2 \mu_B/T}] ,
\label{eq:pbarp}
\end{equation}
is entirely quantified by the $\mu_B/T$ value. Thus,  an  increase
in the \pbar/p ratio from heavy-ion to \pp collisions, seen in
Fig.~\ref{pbarp},      is due to a decrease in the $\mu_B/T$
value.

To extract the corresponding $\mu_B$ and $T$  at fixed energy  in
\pp collisions we have used the THERMUS
code~\cite{Wheaton:2004vg,Wheaton:2004qb} which correctly accounts for feeding
corrections to (anti)nucleons from decays of heavier resonances.

The \pbar/p ratios measured in \pp collisions, shown in
Fig.~\ref{pbarp}, have been fitted using the statistical thermal
model by varying only the parameters $d$ and $e$ in
Eqs.~(\ref{Eqn:T}) and (\ref{Eqn:mub}). We have used the same
$T(\sqrt{s_{\rm NN}})$ dependence for \pp as for heavy-ion
collisions. This is justified by the observation that at high
energies there is no noticeable change in $T$ between central and
peripheral  heavy-ion as well as \pp collisions
\cite{Kraus:2007hf}.
The resulting  baryo-chemical potential $\mu_B$ is shown  in the
lower part of Fig.~\ref{pbarp} by filled circles. In addition,
applying the parametrization of $\mu_B(\sqrt{s_{\rm NN}})$ as in
Eq.~(\ref{Eqn:mub}) we have found that the parameters
corresponding to \pp collisions are
%give more prominence to the new analysis of Natasha
% and change the notation to avoid confusion with the HI parametrization
\begin{equation}
%\mathrm{p-p~~Collisions:~~}~~
\mu_B = d_{\rm pp}/(1+e_{\rm pp}\sqrt{s_{\rm NN}})
\end{equation}
 with $d_{\rm pp} = 0.4$ GeV and $e_{\rm pp} = 0.1599$ GeV$^{-1}$.
 The solid line in the  lower part of
Fig.~\ref{pbarp} represents the energy dependence of $\mu_B$ in
\pp collisions obtained with the above parameters. For comparison
also shown in this figure is the energy dependence of the value of
$\mu_B$ in heavy-ion collisions.  From  Fig.~\ref{pbarp} it is
clear  that at mid-rapidity,  the  {\muB} is  always lower in \pp
than in heavy-ion collisions. This observation reflect the fact
that at mid-rapidity the stopping power in \pp collisions is less
than in heavy-ion reactions. The change of \pbar/p ratio with
energy in \pp collisions   is quantified in the upper part of
Fig.~\ref{pbarp} using  parametrization of $\mu_B(\sqrt{s_{\rm
NN}})$ adjusted for \pp collisions.

\begin{figure}
\includegraphics[width=0.99\linewidth]{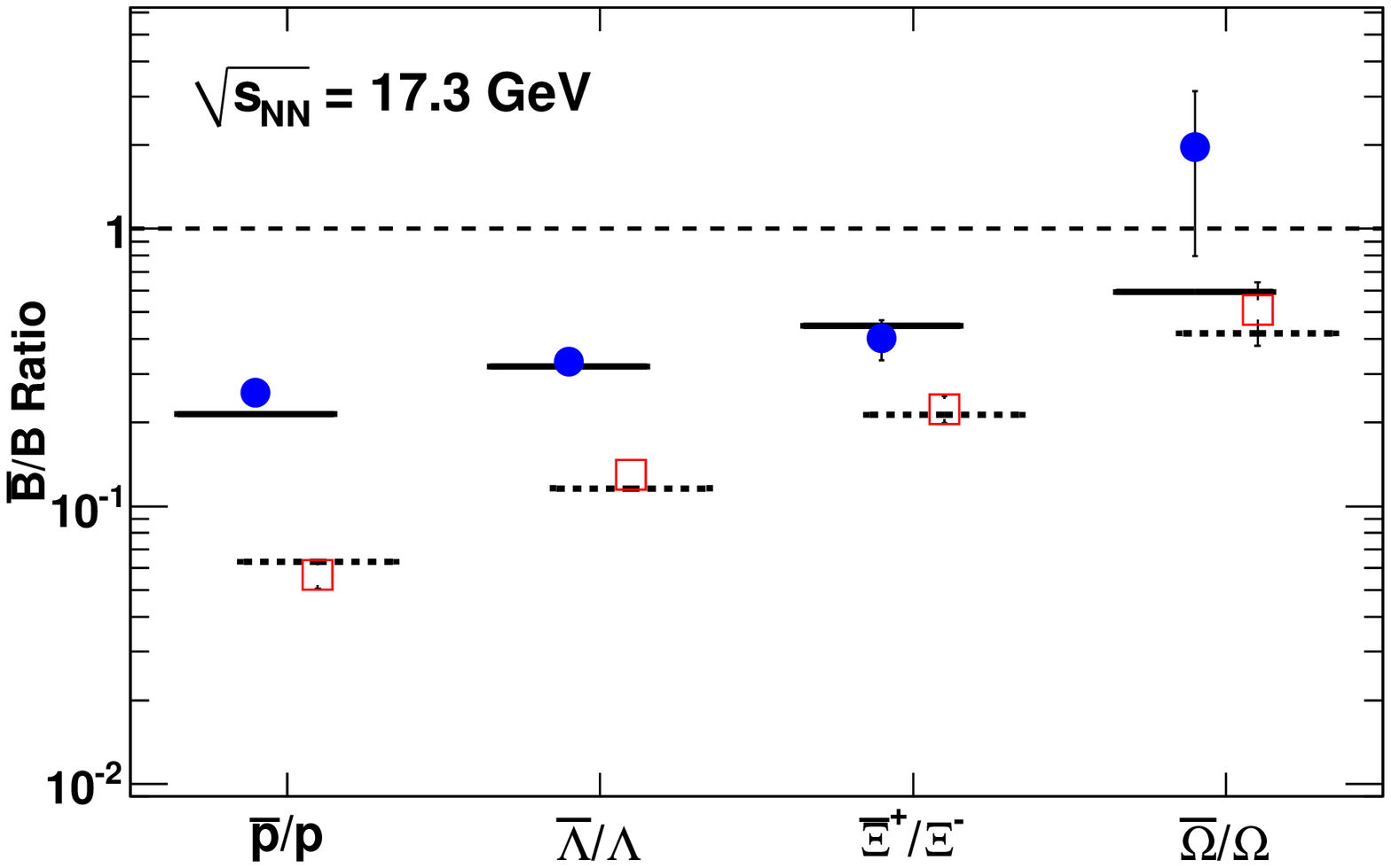}
\includegraphics[width=0.99\linewidth]{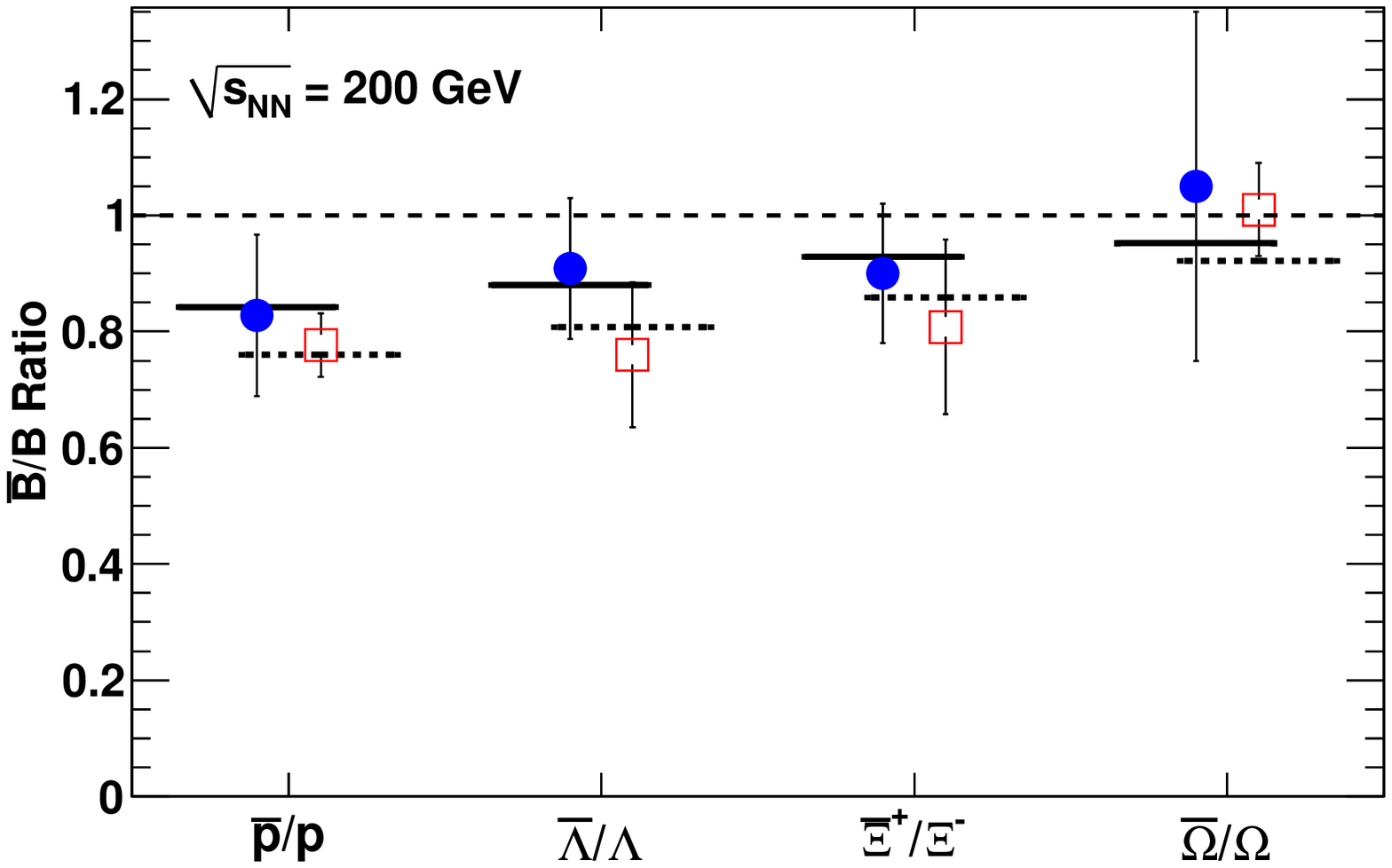}
\caption{Antibaryon to baryon ratios sorted according to their
strangeness content. Circles (solid horizontal line) refer to \pp
collisions data (model calculations) and open squares (dashed
horizontal line) refer to heavy-ion collisions data (model
calculations). The upper part shows results at the SPS and the
lower part at the RHIC energy.} \label{BbarB_ratio}
\end{figure}
For baryons carrying a number of $N_S$ (anti)strange  quarks the
antibaryon/baryon ratio (again neglecting for the moment feed-down
from resonances):
\begin{equation}
\frac {n_{\rm\overline{B}}}{n_{\rm B}} =  \exp[{-2( \mu_B - N_S
\mu_S )/T}],  \label{eq:BbarB}
\end{equation}
is modified by the  strange chemical potential. As \muS~is always
smaller than \muB\ (see e.g.~\cite{Redlich:2002ij,Cleymans:1992zc}), the above ratios
should appear ordered with the strangeness quantum numbers,
i.e.~the higher $N_S$, the smaller the difference between
antibaryon and baryon.

Figure~\ref{BbarB_ratio} shows data on strange antibaryon/baryon
ratios  from the SPS and RHIC energies and their comparisons  with
the model calculations using the THERMUS code. The data and the
model results both in \pp and heavy-ion collisions  are in good
agreement. There are   clear trends in strange antibaryon/baryon
ratios already expected from the simplified Eq.~(\ref{eq:BbarB}).
(i) With increasing strangeness quantum number, the
antibaryon/baryon ratios are increasing and approaching to unity.
(ii) Heavy-ion collisions exhibit smaller \Bbar/B ratios as
compared to \pp collisions due to different \muB\ values as
mentioned before. This is well seen at SPS energies, where the
difference in \muB\ in \pp and Pb+Pb systems is larger than at
RHIC. (iii) The differences between heavy-ion  and \pp collisions
decrease with increasing \sNN. At LHC energies the \pbar/p ratio
is close to  unity  and therefore, the abundances  of strange
baryons are roughly as large as
%added "those of"
those of  their antiparticles.

\section{\label{secAnalysis} Production of (anti)nuclei and (anti)hypernuclei}

\subsection{Comparison to data from RHIC}

The production  of  antimatter compared to matter,  being
expressed by ratios of {anti}baryon/baryon yields,  was shown in
the last section to be well described by the statistical thermal
model. Thus, it is of interest to verify whether the recently
observed production of light (anti)nuclei including
(anti)hypertritons ($^3_\Lambda$H)  in heavy-ion collisions at
RHIC by  STAR Collaboration~\cite{Abelev:2010} also  follows a
pattern expected in  the statistical thermal model.

Studying the antinuclei/nuclei ratio in the statistical thermal
model  an extra factor of $\mu_B$ is picked up each time the
baryon number is increased. Thus, each nucleon adds a factor of
$\mu_B$ in the exponent of the Boltzmann factor in
Eq.~(\ref{eq:dens}). The production of nuclear fragments is
therefore very sensitive to the  value of the baryo-chemical
potential and  thus could be used for a precise determination of
$\mu_B$.

The deuterium has  an additional neutron and the
antideuterium/deuterium ratio in the statistical thermal model is
given by
\begin{equation}
\frac {n_{\rm\overline{d}}}{n_{\rm d}}       =  \exp[{-4 \mu_B
/T}], \label{eq:dbard}
\end{equation}
thus should be similar to  the square of the antiproton/proton
ratio if decay contributions  of heavier resonances to nucleon
yields are neglected. The $^3$He has three nucleons and the
corresponding anti-$^3$He/$^3$He ratio is given by
\begin{equation}
\frac {n_{\rm \overline{^3{He}}}}{n_{\rm ^3He}} =  \exp[{-6
\mu_B/T}], \label{eq:3hebar3he}
\end{equation}
which then is $\sim$(\pbar/p)$^3$.

If the nuclei carry strangeness, this leads to  an  extra term
$\mu_S$ and the ratio of antihypertriton/hypertriton reads
\begin{equation}
\frac {n_{\rm \overline{^3_{\Lambda}{H}}}}{n_{\rm ^3_{\Lambda}H}}
= \exp[{-(6 \mu_B - 2 \mu_S )/T}] . \label{hyperbarhyper}
\end{equation}
In mixed ratios, i.e.~using ratios of different nuclei (or
antinuclei),  there appears an extra factor due to different degeneracy  and masses,
e.g.~in the statistical thermal model
\begin{equation}
\frac {n_{\rm ^3_{\Lambda}H}}{n_{\rm ^3He}}  = \frac {g_{\rm
^3_{\Lambda}H}}{g_{\rm ^3He}}\frac {(m_{\rm
^3_{\Lambda}H})^2}{(m_{\rm ^3He})^2} \frac {K_2(m_{\rm
^3_{\Lambda}H}/T)}{K_2(m_{\rm ^3He}/T)} \exp[ -\mu_S /T] .
\label{hyper3he}
\end{equation}
%with $d_{\rm ^3_{\Lambda}H}/d_{\rm ^3He}=3$.

\begin{figure}
\includegraphics[width=0.99\linewidth]{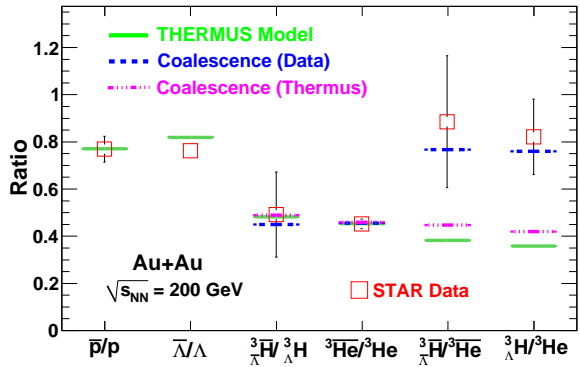}
\caption{Comparison of results from the STAR Collaboration with
the statistical thermal and the coalescence model. For the latter
both experimental values~\cite{Abelev:2006jr,Adams:2006ke} (dashed
lines) and values from the statistical thermal model (dash-dotted
lines) have been used.} \label{STAR_models}
\end{figure}

Figure~\ref{STAR_models} shows comparisons of the statistical
thermal model results on different (anti)nuclei ratios with recent
experimental data from  the STAR Collaboration. These $^3$He and
$\overline{^3{\rm He}}$ yields have been corrected for
contamination  from hypertriton and antihypertriton decays
assuming a decay branch ratio of 25\% and consequently in the
model such decays have not been included.

In the statistical thermal model, following
Eqs.~(\ref{eq:3hebar3he}) and (\ref{hyperbarhyper}),   ratios of
(anti)nuclei/nuclei are entirely quantified  by the $\mu_B/T$ and
$\mu_S/T$ values. From Fig.~\ref{STAR_models} it clear that using
the thermal parameters at chemical freeze-out obtained from the
analysis of particle yields at RHIC, there is an excellent
description of measured ratios of ${\rm \overline{^3He}}/{\rm
^3He}$ and ${\rm \overline{^3_\Lambda{H}}}/{\rm ^3_{\Lambda}H}$.
However, deviations are seen on the level of mixed ratios, ${\rm
\overline{^3_\Lambda{H}}}/{\rm \overline{ ^3He}}$ and ${\rm
^3_{\Lambda}H}/{\rm ^3{He}}$.
%, which nevertheless are not
%exceeding standard deviations.

In elementary collisions  nuclei and antinuclei as well as
hypernuclei and antihypernuclei can be produced by direct pair
production. In heavy-ion collisions, due to final state
correlations, a different  production mechanism  opens up  through
hadron coalescence. Indeed,  production of nuclei  in Pb+Pb
collisions at \sNN = 17.3 GeV at CERN SPS \cite{Arsenescu:2003eg}
have been found to be  consistent with a coalescence picture,
while this was not the case in p+Be collisions at the same energy.

In the most straightforward coalescence picture the ratios of
different (anti)nuclei can be directly related to ratios of
hadronic yields.  In particular,
\begin{equation}
{\rm \frac {\overline{^3 He} }  {^3He } = \frac {\rm \overline{p}
\overline{p} \overline{n} } {\rm  p p n } \simeq ( \frac {
\overline{p} } {  p })^3
 \label{c_3he}
}
\end{equation}

\begin{equation}
{\rm \frac {\overline{^3_{\Lambda}{H}}}
      {^3_{\Lambda}H}=
\frac { \overline{p} \overline{n} \overline{\Lambda} }
      {  p n \Lambda }
\simeq
% \frac { \overline{p}^2 \overline{\Lambda} } {  p^2 \Lambda }
( \frac { \overline{p}  } {  p } )^2
\frac { \overline{\Lambda} } { \Lambda }
      \label{c_hypert}
}
\end{equation}

\begin{equation}
{\rm
\frac { ^3_{\Lambda}H }
      { ^3He}  =
\frac { p n \Lambda }
      {  p p n  }
%\simeq \frac { {p}^2 \Lambda }
%      {  p^3 }
\simeq \frac { {\Lambda} }
      {  p }
      \label{c_hypert_3he}
}
\end{equation}

and

\begin{equation}
{\rm \frac { \overline{^3_{\Lambda} {H}   }}
    {\rm  \overline{^3 He} }
=
\frac { \overline{p} \overline{n} \overline{\Lambda} }
      {  \overline{p} \overline{p} \overline{n} }
%\simeq \frac { \overline{p}^2 \overline{\Lambda} }
%      {  \overline{p}^3 }
\simeq \frac {  \overline{\Lambda} }
      {  \overline{p} }   .
\label{c_anti_hypert_3he}
}
\end{equation}

Comparing results of  the statistical thermal model with the
coalescence framework, one sees that some ratios are quite
similar. Indeed, from  Eqs.~(\ref{eq:3hebar3he}) and (\ref{c_3he})
as well as from Eqs.~({\ref{hyperbarhyper}) and (\ref{c_hypert})
it is clear that neglecting feed-down from resonance decays the
statistical thermal model coincides with  coalescence predictions
on the level of ${\rm \overline{^3{He}}}/{\rm ^3He}$ and ${\rm
\overline {^3_\Lambda{H}}}/{\rm ^3_{\Lambda}H}$ ratios (see also
Ref.~\cite{Andronic:2010qu}). Thus, as long as the key input
ratios \pbar/p and $\bar\Lambda/\Lambda$ are in agreement with a
thermal descriptions, the measured ratios do not allow to
distinguish the two mechanisms. However, differences between these
models are seen on the level of mixed ratios, ${\rm
\overline{^3_\Lambda{H}}}/{\rm \overline{^3{He}}}$ and ${\rm
^3_{\Lambda}H}/{\rm ^3{He}}$, due to different masses of nuclei.
From Eqs.~(\ref{hyper3he}) and (\ref{c_hypert_3he}) one finds that
when neglecting binding energy of nuclei and feed-down corrections
the statistical thermal model differs from the coalescence
framework by a factor of $(1/3+2{\rm m_p}/3{\rm
m}_\Lambda)^{3/2}$. Consequently, the statistical thermal model
results for ${\rm \overline{^3_\Lambda{H}}}/{\rm
\overline{^3{He}}}$ and ${\rm ^3_{\Lambda}H}/{\rm ^3{He}}$ ratios
(solid lines in Fig.~\ref{STAR_models}) are lower than those
obtained in the coalescence picture using the (anti)\La/p ratios
from THERMUS.

The  results from the coalescence
model~\cite{Sato:1981ez,Ioffe:2003uf}  are compared to data from
the STAR Collaboration  and the statistical thermal model
predictions in Fig.~\ref{STAR_models}. The coalescence estimate
has been done using  the  \pbar/p, $\bar{\Lambda}/\Lambda$,
\Lbar/\pbar\ and \La/p  ratios both measured by the STAR
Collaboration~\cite{Adams:2006ke,Abelev:2006jr,Abelev:2010}
(dashed lines) and from the THERMUS calculations (dash-dotted
lines).

We note that  in coalescence picture the equilibrium abundances of
particle yields are not required. Consequently, (anti)nuclei
produced  from  the off-equilibrium medium can   lead to particle
ratios being  in agreement with the simple coalescence estimate
discussed above. However, this is not anymore the case for
statistical thermal model which requires statistical order of
particle yields in the final state.

\subsection{Predictions for RHIC and LHC}
In the previous section we  concentrated on the  statistical
thermal model description of (anti)matter production in heavy-ion
collisions  up to  RHIC energies. In the following we extend our
discussion to higher incident energies and   quantify differences
between \pp and heavy-ion collisions.

In Fig.~\ref{AuAu_pp_200_7} we compare \pp and heavy-ion
collisions at \sNN\ = 200 GeV. In the context of the statistical
thermal model the difference between these two colliding systems
is caused by different values of  $\mu_B$   and by the effect of
canonical suppression in \pp collisions. The ratios of
antinuclei-to-nuclei without strangeness content are only affected
by the baryo-chemical potential which at mid-rapidity is smaller
in \pp than in heavy-ion collisions as discussed earlier.

Figure~\ref{AuAu_pp_200_7}, upper part,  nicely demonstrates that
with increasing mass  the effect of \muB\ becomes stronger, yet, a
strangeness content causes an opposite trend as discussed earlier.
The ratio of hypertriton-to-$^3{\rm He}$ and the corresponding
antimatter ratio show the effect of the canonical suppression
reducing the yield of (anti)baryons carrying strangeness. For the
chosen correlation volume with $R_c = 1.5$ fm the difference is
not dramatic but very noticeable.
\begin{figure}
\includegraphics[width=0.99\linewidth]{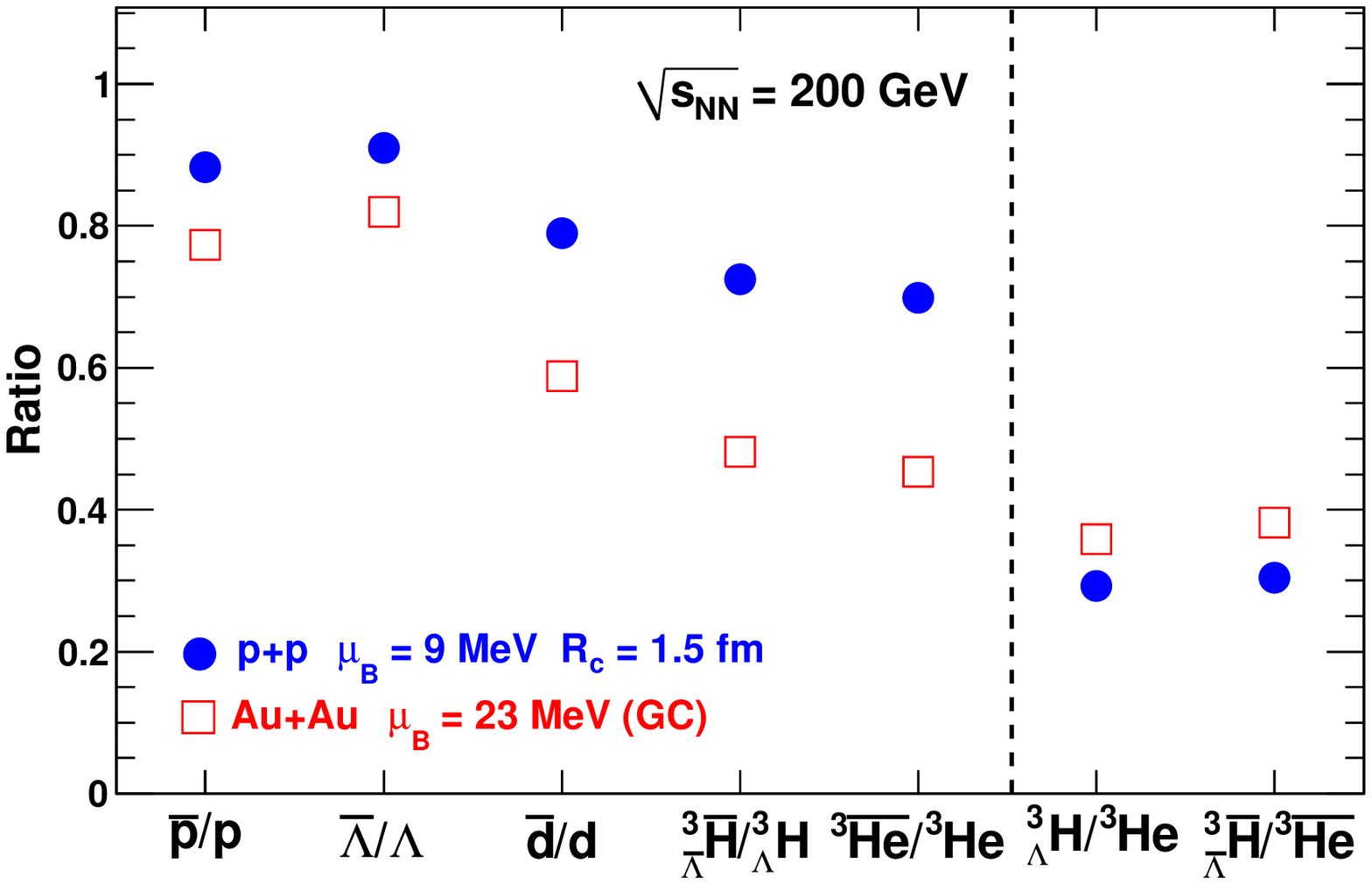}
\includegraphics[width=0.99\linewidth]{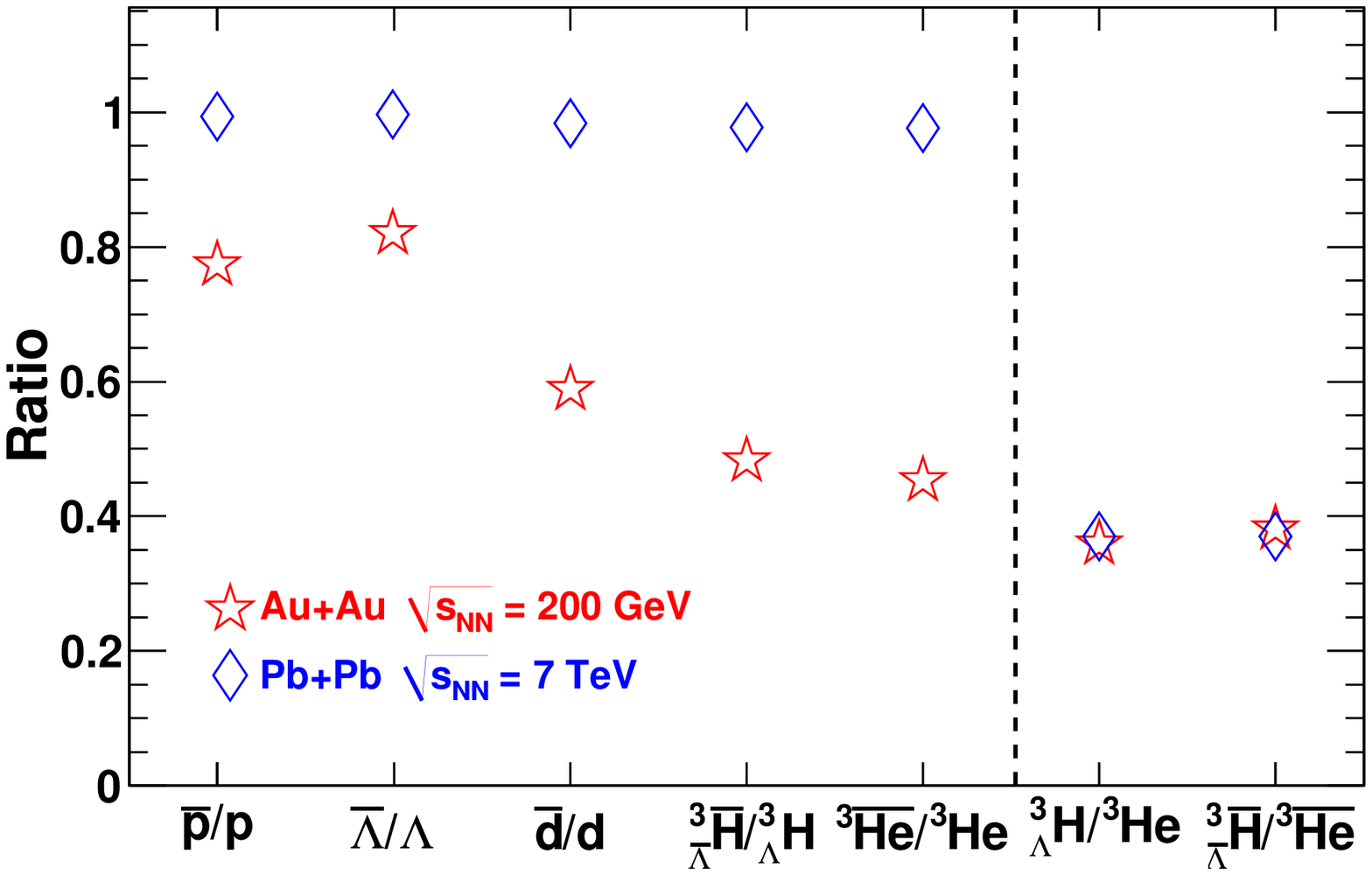}
\includegraphics[width=0.99\linewidth]{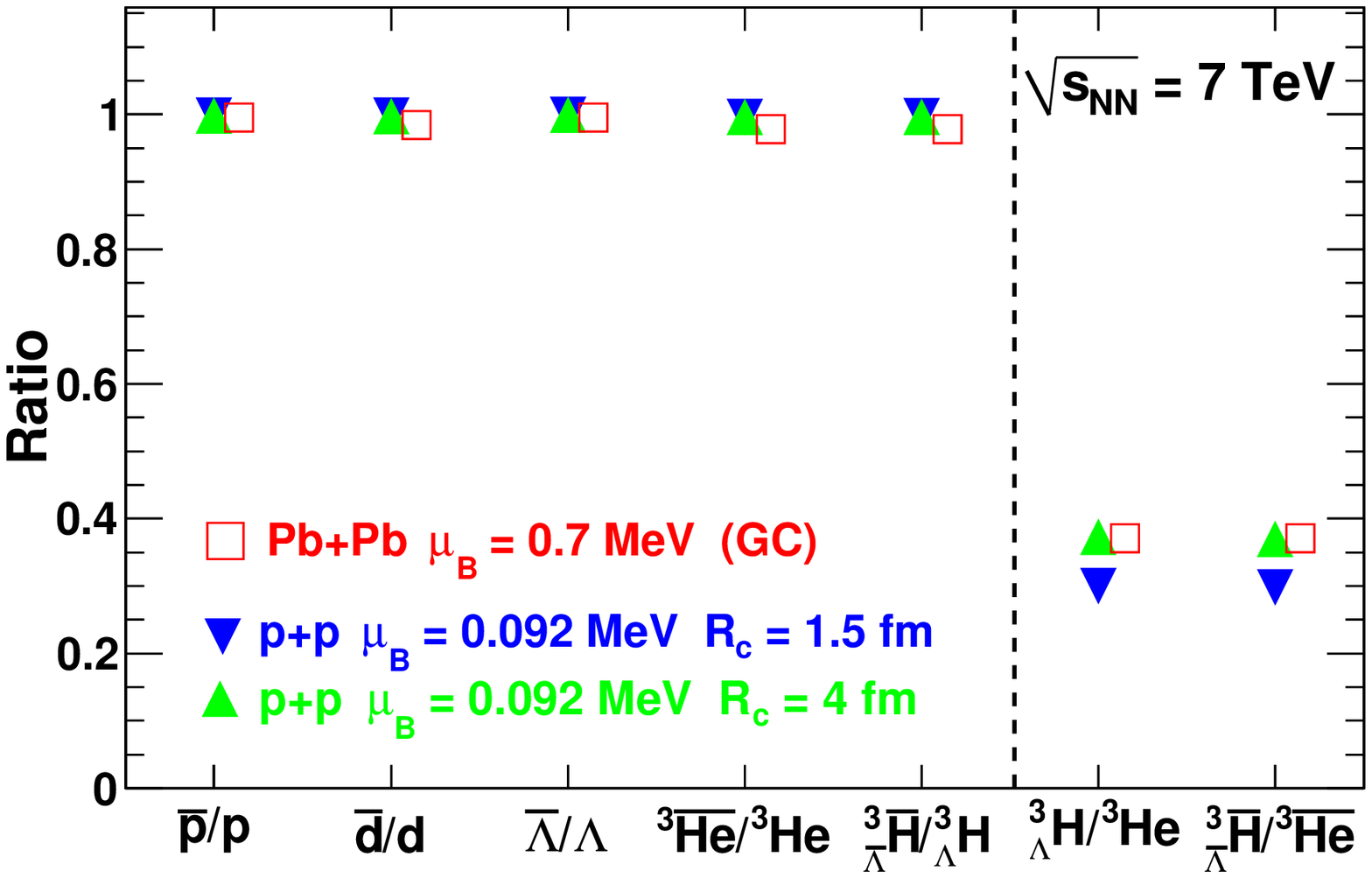}
\caption{Comparison of different  particle ratios calculated in
the statistical thermal model using $T$ = 170 MeV. Upper figure:
for \pp and heavy-ion collisions at \sNN\ = 200 GeV. Middle
figure: For heavy-ion collisions at different collision energies.
Lower figure: Prediction for \pp and Pb+Pb collisions at \sNN\ = 7
TeV.} \label{AuAu_pp_200_7}
\end{figure}

The effect of increasing collision energy is demonstrated in the
middle part of Fig.~\ref{AuAu_pp_200_7}. Here, the differences
between  the antimatter/matter ratios in heavy-ion collisions at
RHIC and LHC are essentially due to the  decreasing value of
$\mu_B$. At LHC,   the chemical potential is smaller than 1 MeV
resulting in the antimatter/matter ratio being close to unity. The
ratios of the (anti)hypernuclei/$^3$(anti)He remain nearly
unchanged from RHIC to LHC since here the effect of $\mu_B$ is
only due to the strange chemical potential which is small. These
ratios are dominated by mass differences and  degeneracy factors.

The expectations for LHC energies are studied in more detail in
the lower part  of  Fig.~\ref{AuAu_pp_200_7}, by comparing the \pp
and Pb+Pb collisions. For simplicity, in both cases the collision
energy of 7 TeV has been chosen. The ratios do not change between
\sNN\ = 2.76 TeV and 7 TeV. The antimatter/matter ratios are
hardly changed from \pp to heavy-ion collisions. All
antiparticle/particle ratios are close to unity. The ratios of
(anti)hypernuclei/$^3$(anti)He exhibit the influence of the
canonical suppression for the correlation volume (see
Section~\ref{model}) corresponding to $R_c=1.5$
fm~\cite{Hamieh:2000tk}. For larger $R_c$ the canonical effect is
reduced and already for $R_c=4$ fm is hardly visible.

The predictions of the statistical thermal model for ratios of
particles with different masses are shown in Fig.~\ref{mass_diff}.
The calculations have been performed for Au+Au collisions at \sNN\
= 200 GeV.

For the results presented in the preceding figures, we used the
freeze-out temperature according to Eqs.~(\ref{Eqn:T}) and
(\ref{Eqn:mub}). It is clear that ratios of nuclei with different
masses are strongly influenced by the value of the freeze-out
temperature. Figure~\ref{mass_diff} displays statistical thermal
model results obtained with  $T$ varying between 110 MeV and 170
MeV. These calculations are compared with the recently measured
value from the STAR Collaboration including the observation of
anti-alpha particles~\cite{Agakishiev:2011ib,Adler:2001uy}. More
data are needed before a freeze-out temperature for antinuclei can
be concluded.

\begin{figure}[hbt]
\includegraphics[width=0.99\linewidth]{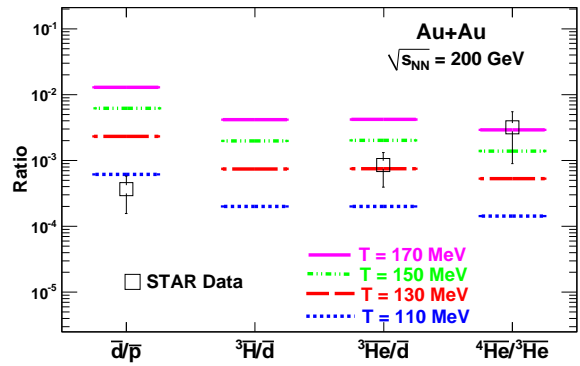}
\caption{Comparison of model calculations of various nuclei ratios
with different masses for Au+Au collisions at \sNN\ = 200 GeV for
different freeze-out temperatures in comparison with the recently
measured values~\cite{Agakishiev:2011ib,Adler:2001uy}.}
\label{mass_diff}
\end{figure}

\section{\label{secSummary} Summary}

We have discussed in a systematic manner  the production  of
(anti)matter in high energy collisions within the statistical
thermal model. We have focused on a general comparison of the
production of antibaryons and antinuclei. The variation of the
\pbar/p ratios with \sNN\ being different for \pp and heavy-ion
collisions has been used to obtain the   parametrization of the
energy dependence of thermal parameters  in \pp collisions beyond
the SPS energy. We have demonstrated the  scaling behavior of the
(anti)baryon/baryon ratios with the strangeness quantum number and
the changes  in these ratios    between \pp and heavy-ion
collisions  with \sNN.

We have compared the measured  ratios of nuclear and anti-nuclear
fragments in  heavy-ion collisions with the statistical thermal
model and with the coalescence concept. Based on the successful
description of existing data, we have made  predictions for
(anti)matter production in \pp and heavy-ion collisions at LHC
energies.

\begin{acknowledgments}
We acknowledge  supports of the Deutsche Forschungsgemeinschaft  (DFG), the Polish  Ministry of Science
(MEN),  the Alexander von Humboldt Foundation and
the ExtreMe Matter Institute (EMMI). The financial
supports of the BMBF, the DFG-NRF, the Department of Science and
Technology of the Government of India and the South Africa -
Poland scientific collaborations are also gratefully acknowledged.
\end{acknowledgments}

\bibliography{anti_bib}
%\begin{thebibliography}{99}
\end{document}